**Title**

# Toward path-invariant embeddings for local distance source characterization

**Authors**

Lisa Linville[1], Chengping Chai[2], Nathan Marthindale[2], Jacob Smith[2], Scott Stewart[2], Asmeret Naugle[1]

Sandia National Laboratory
1515 Eubank Blvd. SE
MS 0401
Albuquerque, NM 87123
llinvil@sandia.gov

1.Sandia National Laboratories, Albuquerque, New Mexico 87123, USA

2.Oak Ridge National Laboratory, Oak Ridge, Tennessee 37830, USA

## Abstract

This work builds on recent advances in foundation models in the language and image domains to explore similar approaches for seismic source characterization. We rely on an architecture called Barlow Twins, borrowed from an understanding of the human visual cortical system and originally envisioned for the image domain and adapt it for learning path invariance in seismic event time series. Our model improves the performance on event characterization tasks such as source discrimination across catalogs by 10-12% and provides more reliable predictive uncertainty estimates. We suggest that dataset scale and diversity more than architecture may determine aspects of the current ceiling on performance. We leverage decision trees, linear models, and visualization to understanding the dependencies in learned representations.

Plain Language Summary

We use regional and global seismic event catalogs to explore feature generation with deep neural networks. We use methods that encourage neural networks to learn source related attributes by considering distortions introduced by travel path as noise and learning invariance to them. Our method improves source classification performance on new events but still encodes bias that we suggest more data may improve.

## Introduction

Data driven models such as deep neural networks (DNN) are an increasingly common modelling approach applied to seismic event processing tasks (Mousavi and Beroza, 2022; Si et al., 2024). DNN models that exploit long temporal waveform sections can be particularly vulnerable to encoding specific environmental and sensing conditions that are expressed in available datasets (e.g. distance from source) which can limit their transferability to new datasets. Even though correlated rather than causal features can lead to poor generalization, supervised DNN models within static collection environments are often highly performant given the inherent repeating nature and geospatial consistency of many seismic sources and receivers. In this work we seek solutions that lead to better performance under more dynamic conditions for single station observations, which is a position often encountered for events of interest in new regions which may not be regularly monitored with open networks.

Transfer learning (Chai et al., 2020) or learning across joint catalogs can be a solution to poor generalization from models trained in different regions, but these approaches scale poorly for dynamic monitoring scenarios, may still not perform without sufficient data, and can have undesirable secondary consequences for uncertainty estimation (Linville et al. 2024). Foundation models (Bommasani et al., 2021), which are typically large models trained across diverse and abundant data using self-supervised training techniques, are increasingly attractive where label-free learning at scale can lead to representations that generalize across tasks and datasets. Here we explore self-supervised training strategies that elevate feature learning beyond correlated observational effects for models that are intended to be used for predicting source related attributes. Examples of common source related attributes for seismic event analysis are magnitudes, depths, locations, or moment tensors.

In this work we focus specifically on event type discrimination and offer observations related to self-supervised learning at modest scales on seismic data. We suggest that scale in memory is not a valid proxy of informativeness when relying on datasets built from event catalogs produced by persistent monitoring operations, even though these are often the most prolific curated data sources in the seismic domain. Self-supervision is what may allow us to exploit data with incomplete or limited labels, but decoupling seismogenic traits from observational and environmental factors through self-supervision is not guaranteed nor should it be expected without adding guardrails during learning that enforce such decoupling. We introduce one potential avenue for building seismic foundation models for source attribute assignments by learning path invariance through self-supervision and demonstrate it on the task of event discrimination. Our approach increases the generalization capability by 10-12% compared to supervised approaches and provides more useful predictive confidence.

## Background

To date, the most common deep learning architectures applied to seismic data are convolutional neural networks trained with supervised learning (see review by Mousavi and Beroza, 2023). While the glut of labeled data has enabled the training of models that surpass the performance of more traditional methods (Guo et al., 2020; Mousavi et al., 2019; Mousavi et al., 2020; Ross et al., 2018; Tibi et al., 2019; Zhou et al., 2019) the transferability and generalization capabilities of supervised models remains limited especially when longer duration segments of a signal are needed as input to maintain high performance (Maguire et al., 2024). Seismic models exploit different temporal segments and segment lengths, depending on the task. Event attribute assignments (event type, magnitude, depth) often benefit from longer duration signals or later arrivals compared to signal detection or polarity analysis. Longer duration input signals (60-90 sec) are highly specific compared to shorter ones and single-station models typically have limited incentive to learn invariance to specific attributes in a signal that are acausal but predictive in training datasets.

Common approaches to learning invariance to inessential information in the input leverage the use of objective functions, architectures, and sampling strategies together to learn better data representations. In image modelling, for example, common practice is to use reconstruction error together with data augmentations such as noise, translations, rotations, and flips to learn invariance to similar perturbations in the wild. Similarly in the seismic domain, invariance to noise (Wang et al., 2024) or window alignment shifting can be valuable, whereas other perturbations typically employed in the image domain like image flipping may not provide practical value for algorithms that ingest temporally sequential signals. Applying noise augmentations to seismic data is one of the most direct translations of discriminative self-supervised learning from the image domain (e.g. Caron et al., 2020; Chen et al., 2020; Grill et al., 2020; Hadsell et al., 2006; He et al., 2019; Misra & van der Maaten, 2019) to the seismic domain when learning on informationally dense event spectrograms. However, random noise does not accurately represent the signal transformations that occur enroute from a source to different sensing sites and adding it should not be expected to result in path invariant representations.

Other proximal fields such as audio encoding with self-supervision as in Wave2Vec (Baevski et al., 2020) provide analogs that may also be considered for seismic self-supervision. Wave2vec for example exploits the abundance of unlabeled data through self-supervision on continuous audio signal and learns representations that are task aligned in ways that unsupervised learning is unable to achieve. Adopting language modelling approaches like BERT (Devlin et al., 2018) which rely on fill-in-the-blank and next-token prediction provide another viable avenue that has been tested in preliminary ways for seismic modelling (Linville, 2023). More recently, vision transformers and pretraining on image patches and other strategies have achieved state of the art in the image domain when tested across many tasks (Dosovitskiy et al., 2020; Doersch et al., 2015; Oquab et al., 2023). However, these audio and language modality approaches often operate in a semantic space that is discretized by words and subwords through tokenization (Kudo and Richardson, 2018; Senrich et al., 2015; Wu et al., 2016) or enforce alignment with metadata though contrastive learning (see Jaiswal et al., 2020 for review on contrastive learning and Si et al., 2024 for seismic implementation relevant to this work). And although it is possible to thoughtfully map these concepts and their upstream and downstream consequences into learning paradigms appropriate for seismic data, substantive work has yet to be done on how different approaches impact the latent spaces that represent our earth system through seismic observations. Here we seek a more direct line for understanding dependencies in representation learning through moderate-scale self-supervision on seismic data.

When downstream tasks are related to event level attributes one vital behavior a model must demonstrate before transferability should be expected is path invariance. Path in this case most critically means waveform propagation pathway, but in our usage will include aspects of a sample that are controlled by the myriad filters a signal produced by a seismogenic source passes through before a digital record is produced. Our self-supervised approach adopts self-supervised ideas from the image domain but reframes the learning objective for station-level path invariance. This pilot study aims to assess how successfully generalizable seismic source representations can be obtained by learning path invariance with a self-supervised objective function called Barlow Twins (BT). The foundational ideas that underpin BT come from exploration of the visual cortical system introduced by Barlow (1961) and were envisioned as a self-supervision technique in the image domain by (Zbontar et al., 2021). BT is an objective function used during model training that forces two altered views of a sample to encode to similar vectors using the cross-correlation matrix between them. By presenting a model with two altered versions of a signal and then encouraging the diagonals of the cross-correlation matrix to sum to 1 while minimizing contributions from off diagonals the model learns invariance to the applied distortions. Re-envisioned for the seismic domain, we consider the augmentation between signals to be conditions (e.g. path effects) that shape the fingerprint of a specific seismic source as it travels from the origin to an observing station. Thus, instead of applying diverse data augmentations to learn invariance, we exploit the observed distortions introduced through wave propagation on a specific path through a medium to a known location or the approximate Green's function encoded upon arrival.

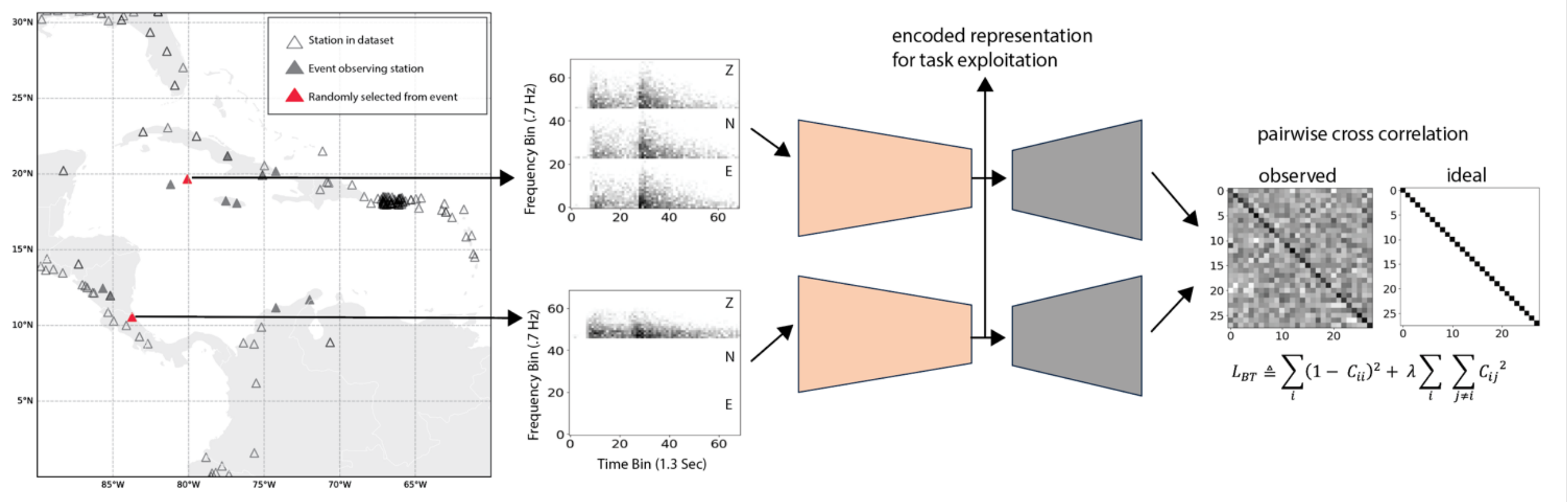

*Figure 1. Model architecture. Pairs of stations are sampled from each event in a batch and the spectrograms are encoded into a 28-dimensional vector through a VGG5 CNN backbone network. The encodings are fed into a projection layer (dim=512). To learn source-centric representations the cross correlation between pairs is penalized the further it is from the identity matrix according to Equation 1.*

Learning source-centric embeddings provides a way to build representations that subsequent fine-tuning for source-focused tasks on limited data can benefit from. For example, Linville et al. (2024) showed that joint training across two distinct catalogs with large geographic coverage does not guarantee or even necessarily benefit performance on geographic subregions contained within them, even with fully labeled datasets. We show that learned representation vectors generalize better to the same regions and that source-focused embeddings provide a natural pathway to accountability that is required for analysts to trust model decisions.

## Data and Method

We exploit several seismic catalogs for model training and testing. For direct comparisons with past work, we use the datasets of Linville et al. (2024) which are derived from several event catalogs from multiple regional contributing seismic monitoring networks across the U.S. (NEIC,

TA, and UUSS in Figure 2). In previous work NEIC earthquake events are gathered for a limited geographic extent and temporal window to study bias. We do not add events to the NEIC catalog to correct the bias, which is primarily the absence of earthquake events west of the U.S. state of Colorado. We then expand to global event catalogs from the United States Geological Survey (Global in Figure 2). We use reported phase arrivals to collect timeseries data from IRIS using FDSN clients available through Obspy (Beyreuther et al, 2010). We use libcomcat to collect reported arrival times. 90 sec duration velocity seismograms are converted to spectrograms (scipy.signal.specgrogram; nperseg = 128, noverlap=0). We use the 3 orthogonal data channels to generate the spectrograms and zero fill if sensors are vertical only receivers. Spectrograms are log scaled and normalized to a range between 0 and 1 for model input. This input representation follows previous convention for seismic event discrimination (Linville et al., 2019; Maguire et al., 2024). We then train our models by exploiting event associations in event catalogs for paired sampling. Event catalogs identify station contributions for each event through event IDs which we randomly sample as pairs for each event available in a training batch (where all stations within an event are maintained in batch sampling). We also perform several additional augmentation steps. First, we add random gaussian noise to each sample to prevent numerical differences from encoding into the embedding and simulate different signal-to-noise ratios. Second, we zero-pad the horizontals from half of the available triaxial station samples per batch and add them as new samples. We pad triaxial stations to develop encodings that leverage but do not require information from all channels, which enables exploitation of a wider range of available sensors. The resulting model is a single station encoder model that exploits event catalogs for their event IDs during learning phases but does not require catalog data or event id for inference or use by subsequent predictive models.

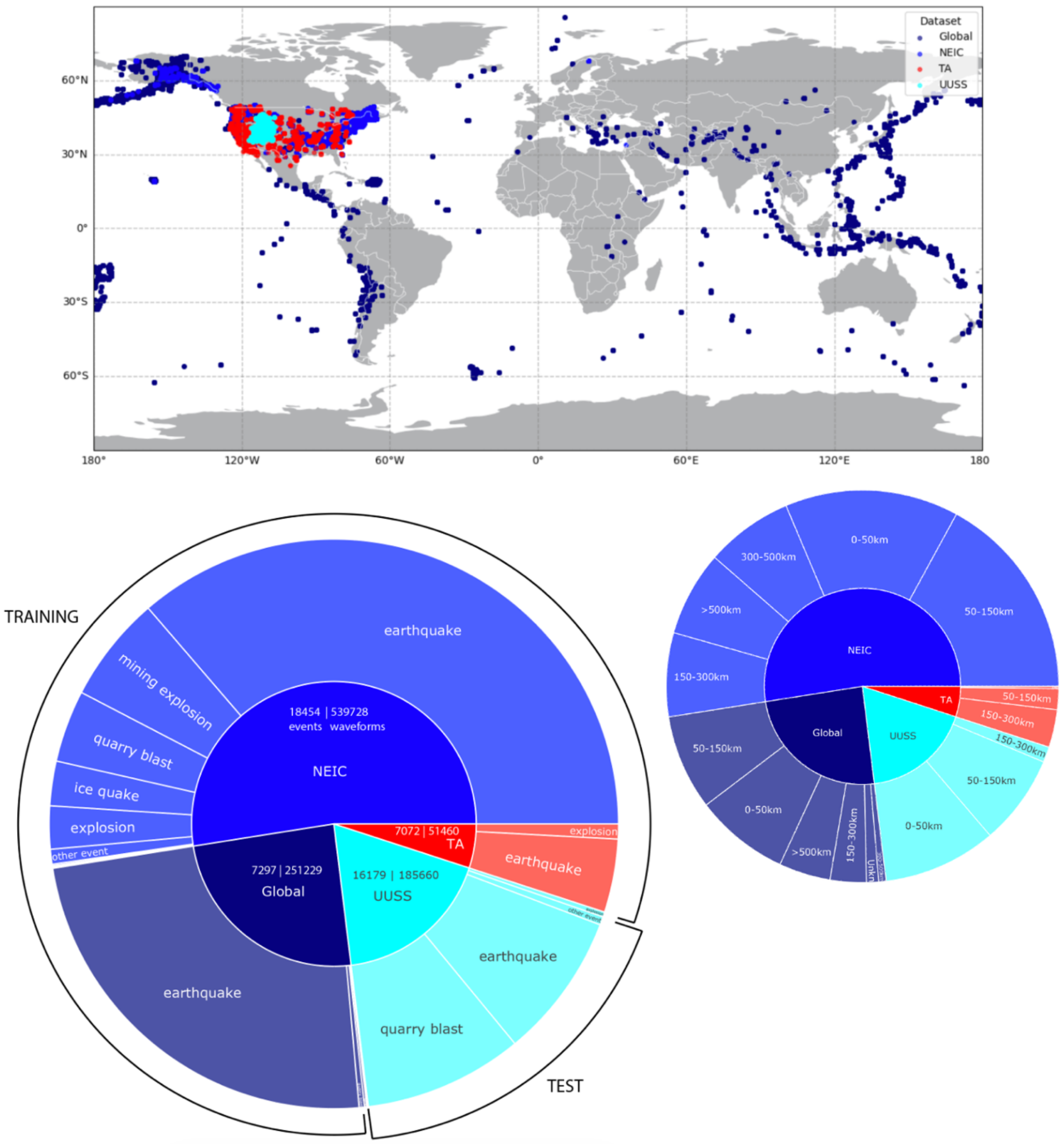

*Figure 2. Datasets used in model training and testing. TA, NEIC, and Global datasets are divided into train (80%), validation (10%), and test (10%) sets randomly partitioned by event ID. Metrics are then reported on a held-out catalog, UUSS. The UUSS events despite their geospatial overlap with training catalogs represent a challenging generalization target reported in supervised learning because of the significant differences in observational characteristics. For each catalog we show the different event type labels (lower left) and the source-receiver distance (lower right).*

We follow the architecture of Zbontar et al. (2021) but use a VGG base model architecture (Simonyan and Zisserman, 2014) to be consistent with past models trained with supervised (Linville et al., 2019) and semi-supervised learning (Linville, 2023). On the VGG base models (VGG5-VGG13) we add a fully connected layer with output size 28 (tested [28, 64, 128, 512]) and projection head size of 512 (tested [2048,4096]). Following the original BT authors, we use the projection layer output for classification tasks, which previous authors call the embedding. We break with their convention using 'embedding' instead for the encoder output layer because it more clearly facilitates relevant comparisons with embedding relationships discussed in the literature. The loss is enacted on the pairwise cross-correlation matrix of the output from the final projection layer for each station pair in a mini-batch (grey in Figure 2). For the loss calculation we directly follow Zbonar et al. (2021):

$$L_{BT} \triangleq \sum_{i}(1 - C_{ii})^2 + \lambda \sum_{i} \sum_{j \neq i} {C_{ij}}^2 \qquad \textit{Equation 1.}$$

Used in this way, BT loss should encourage source attributes to encode on the diagonals (invariance) while minimizing station contributions on the off-diagonals (redundancy reduction). The relative importance of terms is controlled by $\lambda$ , which we found should be small (5E-3) for stable learning as reported by the original authors. Following previous convention, we rely on a VGG backbone, but also test transformer and Variational Auto Encoder backbones (see supplementary section A for details) which offered lower performance. During training, pairs of station observations are sampled across events, which are maintained in each mini-batch. Once trained, the learned representations we use come from the encoder (orange in Figure 2). Models are chosen based on validation accuracy from events excluded from training catalogs. The events in test partitions come from separate catalogs (UUSS; Figure 2) and are distinct from validation and training events. Model selection is discussed in more detail in Supplementary section B.

## Pretraining Analysis

A first order validation of our method is if the event level similarity enforced by the learning objective persists for new events in the test data. Functionally, the effect of learning using Equation 1 is that pairs of samples within training events are forced closer to each other than pairs of samples that are not related. Using a reprojection the 28-dimensional encoder output to a 2D representation (UMAP; McInnes et al., 2018) we visualize the relative locations of samples within events (Figure 3a). The footprint of encoded events does not retain a consistent aperture across observing stations for many reasons but in part because the number of times an example from an event is observed during training depends on the number of observing stations in an event which varies across datasets (TA: 8, NEIC: 23) and within each dataset (most notably by magnitude). We observe that in-event pairs remain closer together than non-event pairs, although in-event apertures for training events are generally smaller than test event apertures (Figure 3a). At the dataset level we show this more concretely by evaluating the cosine similarity of the learned representations (the encoder output) for in-event and non-event pairs sampled randomly across event ID (Figure 3b).

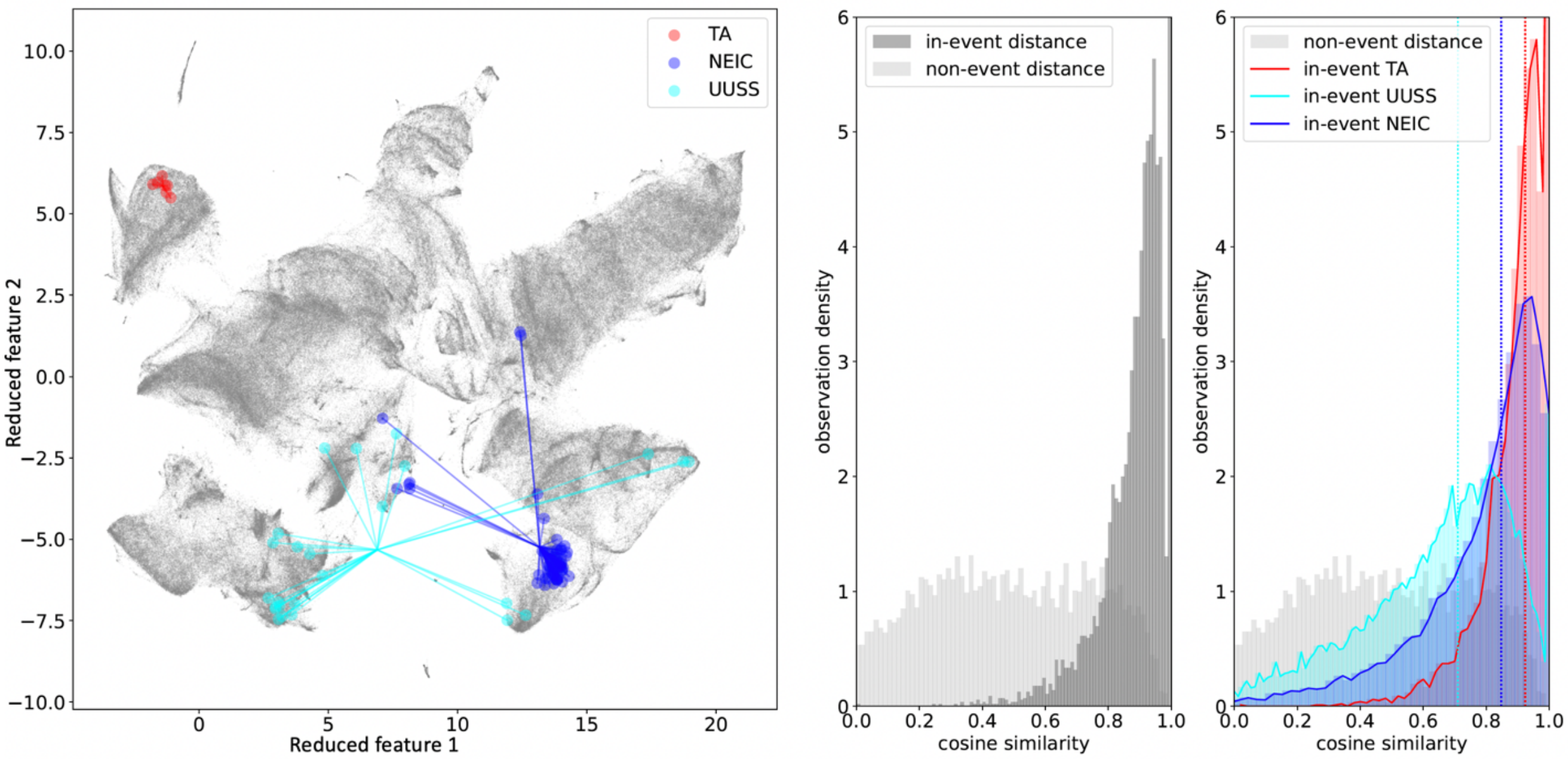

*Figure 3. Distance relationships for encoded samples. Left: UMAP reduced embedding with one event from each catalog randomly chosen. For each event the location of the embeddings for each station are plotted and connected to the centroid. Center: Cosine similarity between in-event and non-event pairs across the training dataset and Right: in-event distances by dataset compared to non-event pairs (gray).*

Distance relationships measured through cosine similarity suggest that enforcing path invariance for training events translates to new events. Evaluated another way, we test how strongly source related attributes from the catalog metadata such as depth and event location rather than path related attributes such as travel distance and station location are predictive of the relationships in learned embeddings reduced through principal component analysis (PCA; Greenacre et al., 2022). We model the importance of metadata attributes on embedding location by looking at feature importance in decision trees trained as predictors of one of the principal components. If an attribute is a strong predictor for a feature (one of the first 2 principal components), we interpret that it plays a meaningful role in the encoded relationships. For comparison we also encode the dataset using a variational auto encoder (VAE; Kingma and Welling, 2013). VAEs use the same bottleneck approach (Tishby & Zaslavsky, 2015; Tishby et al., 2000) that we find effective in the BT architecture but rely on reconstruction of the original input instead of the source-focused objective function. For a baseline comparison we observe metatada relationships for test set predictions (UUSS) using supervised models trained on US catalogs (NEIC and TA) and compare them with UUSS relationships from self-supervised representations on the same catalogs using Barlow Twins. We will later compare performance from representation produced by these same models on the task of event discrimination. We find that path and sensor related attributes like station location and number of available channels are the most important for supervised representations and self-supervised representations with VAE, whereas source related attributes like depth are of primary importance in self-supervised representations with secondary importance from path attributes like distance.

To further verify the predictive value of source-related attributes encoded in the embeddings we relate each feature to a metadata attribute through linear regression and evaluate the coefficients. We find that while no dimensions are entirely descriptive for any single metadata feature, some features are more sensitive to source traits, while others are more sensitive to receiver characteristics. We evaluate the use of these in simple predictive models in the Fine-Tuning section below. Metadata attributes are not independent, and these approaches require parameterization that can have a large impact on the reported feature importance, they therefore do not definitively demonstrate that path invariance is achieved. However, coupled with demonstrated increases in performance they suggest that although path effects still play a role in the learned representations, they are less dominant than in representations learned by supervised or other self-supervised methods. See supplementary section C for more details on linear analysis of embeddings.

## Engendering User Trust: Data Attribution

While the primary objective of this work is to explore approaches that lead to better representations that translate to better performance with minimal data in new regions for source related predictive tasks, we acknowledge that adoption requires trust. To this end we also suggest potential methods for exploring training data attribution. There are many examples in the literature of computationally efficient methods to understand how to identify which specific training examples most meaningfully contribute to how a model interprets a new example (training data attribution, or TDA; Pruthi et al., 2020; Nguyen et al., 2024). In this work we approach TDA through clustering on the UMAP reduced embedding space (from Figure 3) using HSDBScan (Ester et al., 1996; Schubert et al., 2017), which determines the optimal cluster set count based on user parameterization (min_cluster_size=50). We fit the training data (Global and U.S catalogs) and use the fit function to assign categorical cluster assignments for each sample in all datasets. Categorical assignment through clustering provides several advantages: We avoid expensive recursive pairwise cosine similarity calculations, we get a discrete set of values that represent global behaviors that we can the evaluate w.r.t metadata, for example geolocation, and we get information about poorly understood samples or samples that reside in low density areas of the training space based on their cluster membership (-1). In our analysis, performance increased on test data by an additional 11% (U.S. models, 8% for Global models) when using only samples that were assigned a cluster value, suggesting cluster membership as a useful minimal threshold for confident decision making.

We use ipython widgets developed in companion work (Martindale et al., 2024) to explore which samples are most related to test samples based on cluster membership. We find this is a valuable way to relate new predictions to known events in the catalog history. Whether these relationships make sense or are counterintuitive can be useful trust signals for model users (Figure 4). See supplementary section C for deeper exploration of specific global clusters.

In this work we exploit station and event-related metadata only to explore the relationships between learned samples, but also recognize the potential value as a learning signal which we do not use. For example, event latitude and longitude added to the input spectrograms may have a more direct impact on distance relationships in learned embedding that may also generalize and be valuable. Here, we refrain from directly enforcing these relationships to allow for exploration of learned relationships directly attributable to our self-supervised approach. In the future seismic foundation models, as in many large-scale foundation models in the marketplace, may benefit from multimodal data inputs and the use of several learning objectives.

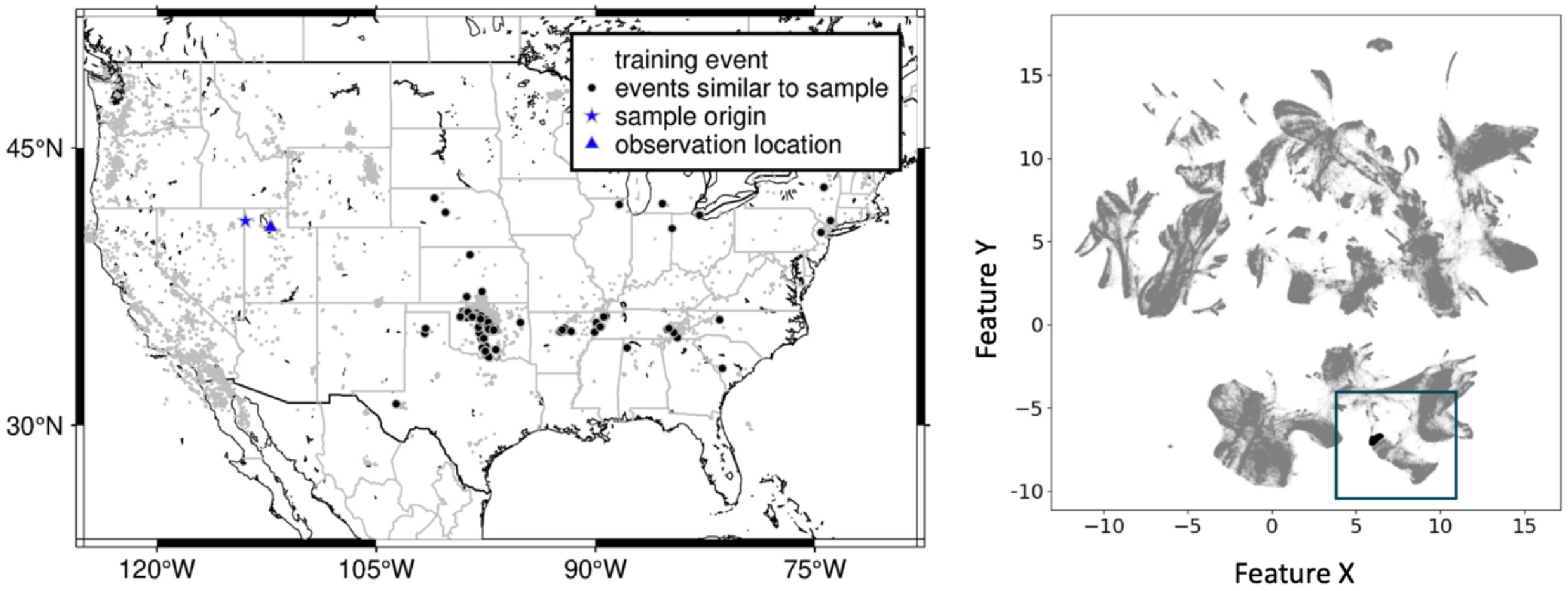

*Figure 4. Map of training data contributions for predictions on new signals. The training data are shown in grey on the map. When a new signal is introduced to the model it gets a cluster assignment which represents the training events that are the most similar (by location in the learned and then reduced embedding space). The embedding reduction for one model is shown on the right with the cluster assigned to the new event shown in black. The corresponding events that belong to that cluster are in black on the map, including the new signal which is plotted in blue by origin (star) and sensor location (triangle).*

## Fine-Tune/Task Performance

So far, we have described exploration of learned embeddings. We now describe the exploitation of the learned embeddings for downstream tasks. We know event discrimination can be a challenging task based on past studies on similar datasets, where test set performance > 95% can drop below 70% on new datasets from new regions (Linville et al. 2024; Si et al., 2024). We compare performance using class balanced accuracy from BT models with performance from supervised learning with similar CNN model architectures (using VGG13 for supervised models and VGG5 for BT models) and a classifier head of the same capacity (4096 parameters) in Figure 5. Figure 5 shows a performance increase from ~ 64% balanced accuracy to 70% with BT trained on U.S. data and up to 72% trained further on global data. Previous work reported lower accuracy for test catalogs (60%) with the difference between our reported values and previous likely being that we did not remove events that were reported across catalogs in pretraining phases. When we build simple logistic classifiers using the embedding dimensions we identified with high coefficients for source-related attributes in the pretraining analysis section, which reduces the representation dimension from 28 to 3, and simplify the classifier from a feed-forward neural network to a logistic regressor trained and tested on the same partitions as above, we achieve very similar performance as BT models (72.3% accuracy).

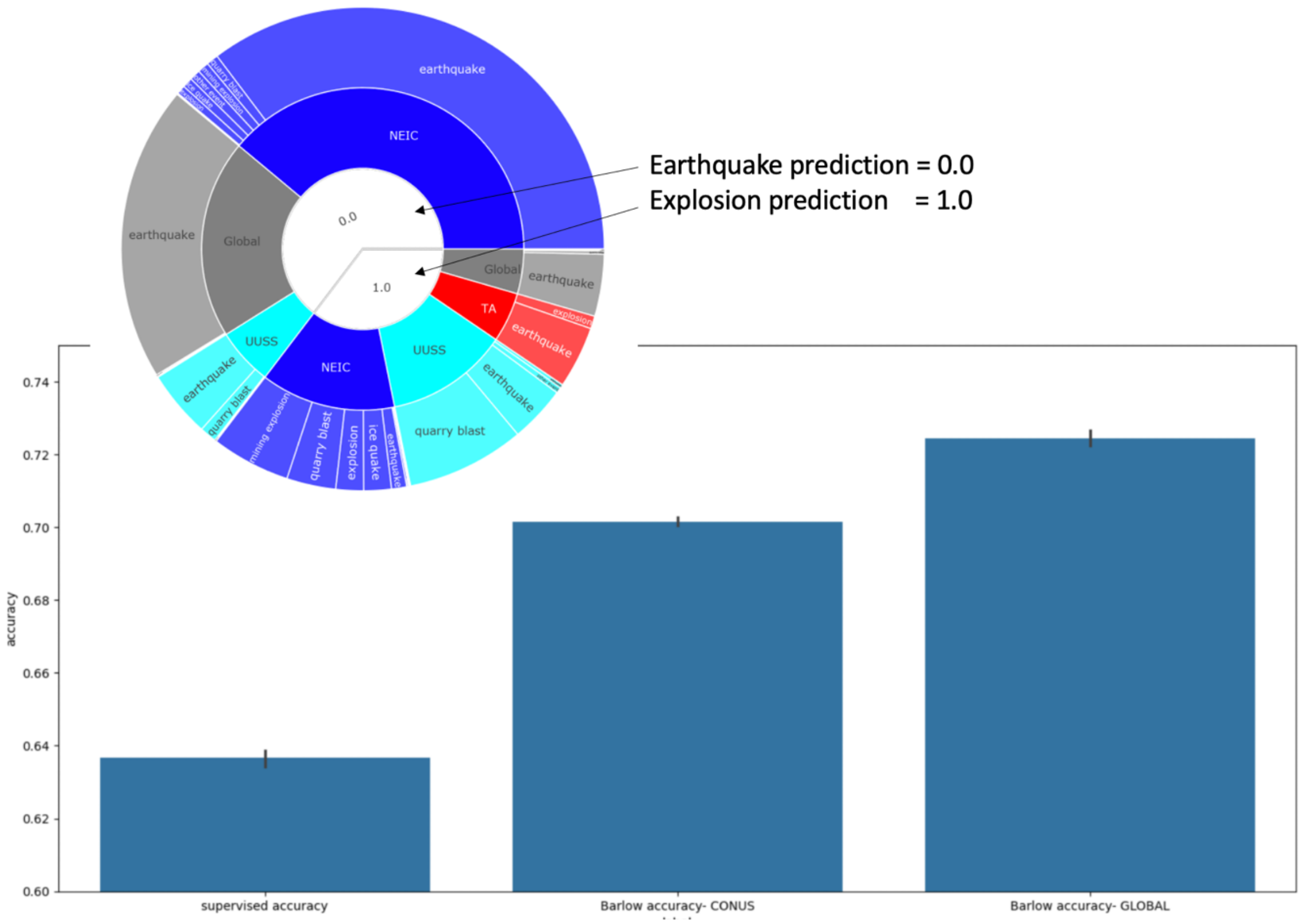

*Figure 5. Accuracy for models trained on different data compared to supervised model accuracy. Training with BT increases accuracy more than adding available informative data.*

One aspect of successful foundation models is that they enable competent learning for new problems with sparse data (Bommasani et al., 2021). The test for this in our application is if training on a few labels from the test catalog results in larger increases in the performance for self-supervised representations than for supervised transfer learning. This was only the case for our models when we severely restricted the size of the available classifier. With abundant and equal expressivity across supervised and self-supervised classifier heads our performance was the same using 2 labels from new test catalogs; our accuracy on test events increased to 80% and supervised learning in this case was much more efficient (< hour compared to several days). Only when we use small dimension (64) linear classifiers do self-supervised representations outperform supervised transfer learning (still 80% accuracy for BT but supervised accuracy is ~ 70%, near the performance of BT with no labels) suggesting that large fully connected non-linear models do not heavily rely on learned features for performance on this task, but when they do or are forced to, path independent features generalize better.

## Discussion

In this work global BT models show modest gains over U.S. catalogs for performance, and we note that these performance increases are similar to those seen for emerging seismic foundation models trained using other self-supervised approaches (Si et al., 2024). While other models are typically larger, we scale our architectures to minimize model parameters while maximizing performance for simplicity. We expect performance for foundation models to scale with data and model size but did not achieve higher performance on downstream tasks with larger base model sizes, embedding dimensions, or projection head sizes suggesting that further investment in the scale and diversity of datasets may be required to fully leverage the architecture for effective global generalization. And while there are many pathways to further model improvement, adding earthquake events from the western U.S. for example, this study shows that without additional investments in datasets, source-focused objective functions can offer representations that generalize and help simplify predictive models for discrimination tasks.

Another advantage our models show is in the quality of the predictive uncertainty they provide. Prediction confidence for supervised models without ensembling or dropout is typically uninformative since the populations for incorrectly predicted samples show a high degree of overlap with the correct distributions (Figure 6 top left). BT models on U.S. data improve this, showing more separability (Figure 6 top right). These populations separate further still in globally trained BT models with nearly uniform distributions for incorrect samples (Figure 6 bottom).

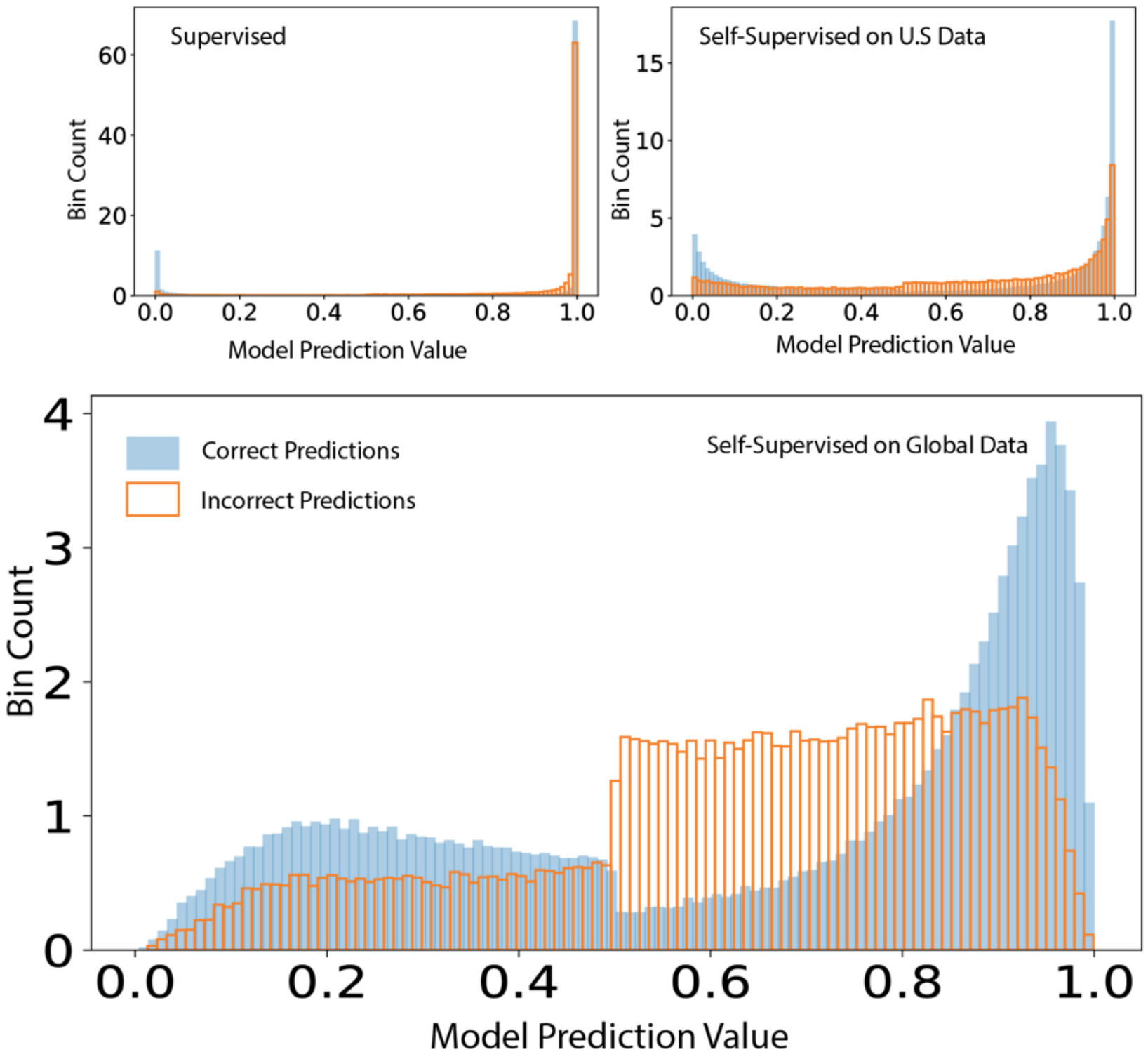

*Figure 6. Predictive uncertainty from BT models. Orange = incorrect and blue= correct decisions. There is a clear bias towards explosive sources (class =1). Confidence values for self-supervised BT models trained on Global data provide the most robust uncertainty values (via the classification output from final model layers. These confidence values are from the UUSS test set.*

We know from previous work that distance and density relationships in embeddings are often semantically descriptive (Micolav, 2013; Ethayarajh, 2019; Dong, et al., 2022). We observe bias in the embedded topology from oversampling within our catalog. For example, instead of a limited number of highly dense clusters we note a disproportionate number of distinct clusters related to events within Oklahoma, which are more closely monitored and more densely measured than some other areas owing to their proximity and potential relationship between wastewater injection in the region (e.g., Hincks et al., 2018). Conversely, earthquakes throughout the western U.S. often fall into a single large cluster. If we measure entropy based on cluster diversity, highly similar events (like two oil field events near Huston, TX) can look more dissimilar that many other highly dissimilar events (an Alaskan volcano event and a mine-site in Utah). Oversampling in our datasets impacts distance and density relationships in the embedding space because of the level of descriptive detail and diversity in those event populations. Models may waste capacity on fitting the space of these observations and introduce bias in the latent space that is learned from the oversampled catalog. Bias and fairness evaluations in the image domain, even with substantial investment in curation expressly to enforce diversity (through deduplication and subsampling) at data scales larger than we explore here still report evidence of bias owing to data sampling (Oquab et al., 2023). In Seismology the issue of developing maximally informative and diverse yet non-redundant data is an open challenge for large scale foundation model learning where emergent properties like world understanding (subsurface interpolation) are often expected but impractical to assume given the distribution of events available in collected catalogs. We mention this specifically as the field advances methods for uncertainty estimation that leverage density relationships (Jiang et al., 2018) and looks toward generative modeling with self-supervision as a pathway to increasing dataset diversity (Eigenshink et al., 2023; Pinaya et al. 2022).

# Conclusions

This work was performed as a pilot study on the viability of discriminative self-supervised learning borrowed from the image domain. The objective was to explore the success of learned invariance to properties of a signal due to propagation pathways and observational conditions that can be damaging to predictive performance. Sensitivity to conditions such as observing distance and other path related attributes can limit the ability to exploit models for source exploration and characterization. We learn path invariance using an objective function called Barlow Twins that when coupled with our data sampling approach encourages a model to learn invariance to path effects and to reduce redundancy across observations. We validate the success of our learning approach by evaluating distance relationships in the learned embeddings and clustering relationships within and across events. We explore the dependencies in learned embeddings with respect to known metadata attributes available in seismic catalogs with decision trees and linear models. We then build simple logistic classifiers using only the source-sensitive dimensions that perform as well as much larger neural network classifiers trained in the fine-tuning phase. We show that orienting the learning around invariance to path effects can result in learned representations that can be effectively exploited in several different ways for downstream tasks related to source characterization.

In conclusion, foundation models are one potential avenue for exploiting large seismic datasets to learn representations that generalize across tasks and regions and reduce the costs associated with training models from scratch when competent models are made available to the broader community. We provide an introduction to one discriminative self-supervised approach with promising results for seismic processing tasks related to assigning source attributes. We also demonstrate room to improve through dataset diversity and model scale. We suggest that for structured seismic tasking, models need only demonstrate successful encoding of aspects of a world model that will be exploited during fine-tuning phases and that doing so can lead to more straightforward evaluation and exploitation of learned embeddings.

## Data and Resources

The UUSS catalog is available for download at https://doi.org/10.31905/RDQW00CT. The UUEB catalog can be downloaded at http://www.sandia.gov/uueb. The NEIC and TA catalogs can be downloaded as CSV tables available in the supplementary section of this paper. The waveform data for all catalogs is available for download from the 'IRIS' FDSN client in the python Obsy library. We use the BT pytorch implementation from facebook research available at https://github.com/facebookresearch/barlowtwins/tree/main implemented with LightningAI using online training as in: https://lightning.ai/docs/pytorch/stable/notebooks/lightning_examples/barlow-twins.html#Barlow-Twins.We rely on PyTorch 3.6 for tensor computation, standard loss and optimization, and automatic differentiation. We use Generic Mapping Tools (GMT) and Seaborn and Matplotlib for figure generation, and Pytorch Lightning parallel for job control. We used MLFlow and Weights and Bias for experiment tracking and support. All model training was performed on multicore GPU clusters maintained by Sandia National Laboratories. Supplemental material for this article is available and includes additional figures and supporting analysis.

## Acknowledgements

This research was funded by the National Nuclear Security Administration, Defense Nuclear Nonproliferation Research and Development (NNSA DNN R&D). The authors acknowledge important interdisciplinary collaboration with scientists and engineers from LANL, LLNL, MSTS, PNNL, and SNL. We recognize our colleagues Dan Krofcheck and Scott Steinmetz at Sandia National Laboratories who contributed technical feedback and early reviews of this work. Sandia National Laboratories is a multimission laboratory managed and operated by National Technology & Engineering Solutions of Sandia, LLC, a wholly owned subsidiary of Honeywell International Inc., for the U.S. Department of Energy's National Nuclear Security Administration under contract DE-NA0003525. This paper describes objective technical results and analysis. Any subjective views or opinions that might be expressed in the paper do not necessarily represent the views of the U.S. Department of Energy or the United States Government.

Supplementary Sections

## Section A: Training Details

In early phases of this work we test Transformer, Variational Autoencoder, and CNN backbone networks to use with Barlow Twin loss. All architectures used pretraining concurrent with online task learning. For each gradient update with pretraining loss we freeze weights, and then run representations through fine-tune task heads backpropagating cross-entropy loss on labels from training sets. Our transformer models did not pass preliminary trials, where we evaluate validation accuracies which did not reach competitive levels compared to CNN models (> 92%). VAE models we did not evaluate w.r.t. task performance but observed encoded representations were highly dependent on observational features such as the number of instrument channels (triaxial vs vertical sensors). See Figure S1 below.

We do not rule out the possibility that transformer architectures may be a good fit for encoding with BT loss and there are many branching ideas that may be valuable to explore with seismic timeseries data and BT loss. We acknowledge that the limitations we observed may be due to insufficient experimentation. Given the stability of learning with CNN base models and the comparisons we make with existing models from the image domain we limit the extent of our exploration for these ideas but invite others to advance our understanding through further experimentation.

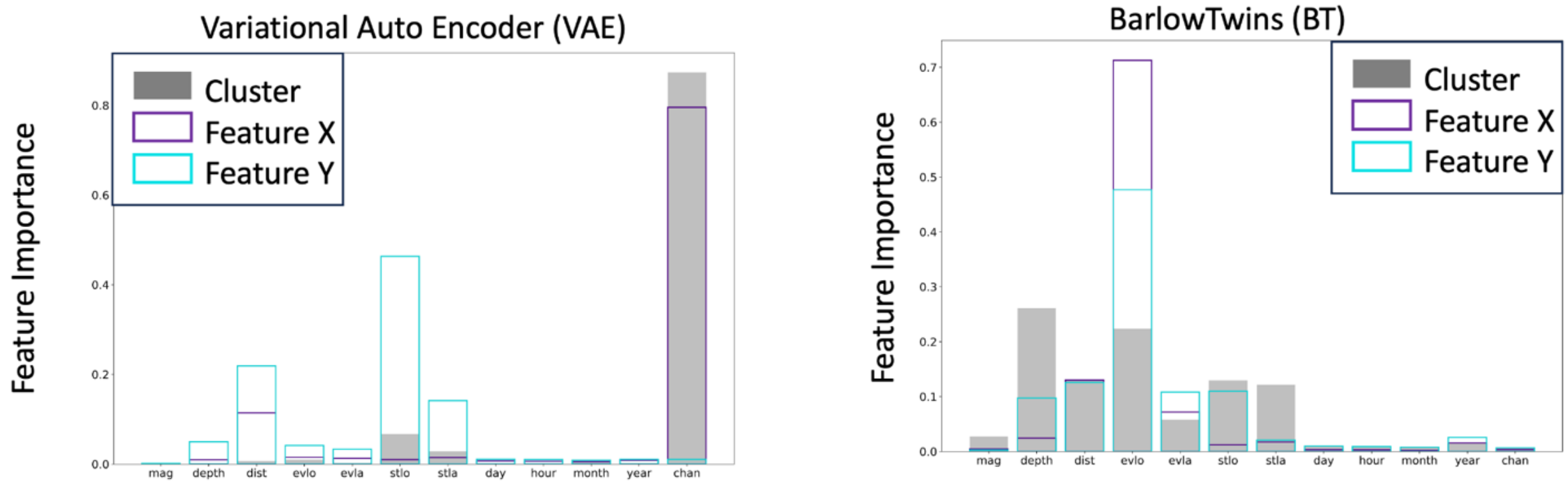

Figure S1. Feature importance using random forest models for VAE and BT models. We test 3 targets (UMAP 2D reduction for feature X and Y and Cluster assignment described in main text) to see how predictive each metadata feature is for each target. VAE models show high dependence on sensor channel dimension (chan) while BT models are more sensitive to source attributes like depth and event longitude.

## Section B. Model selection

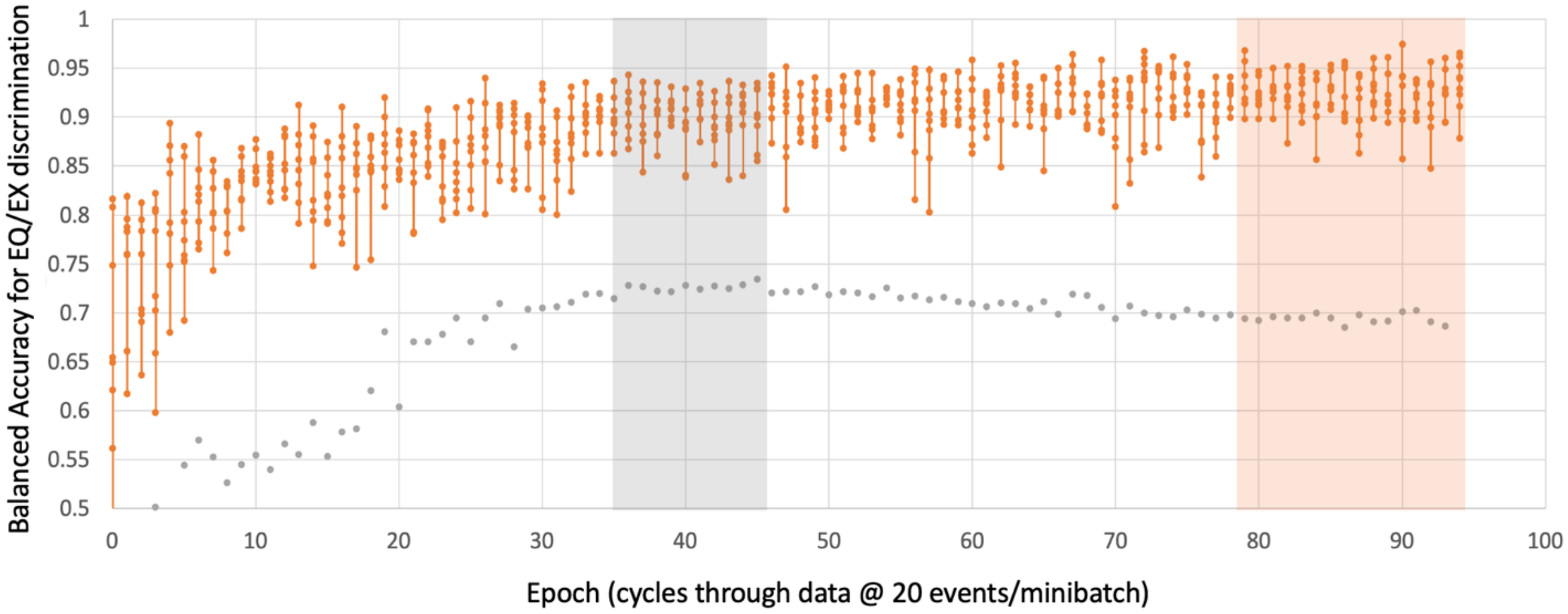
Balanced Accuracy for EQ/EX discrimination
1
0.95
0.9
0.85
0.8
0.75
0.7
0.65
0.6
0.55
0.5
0
10
20
30
40
50
60
70
80
90
100
Epoch (cycles through data @ 20 events/minibatch)

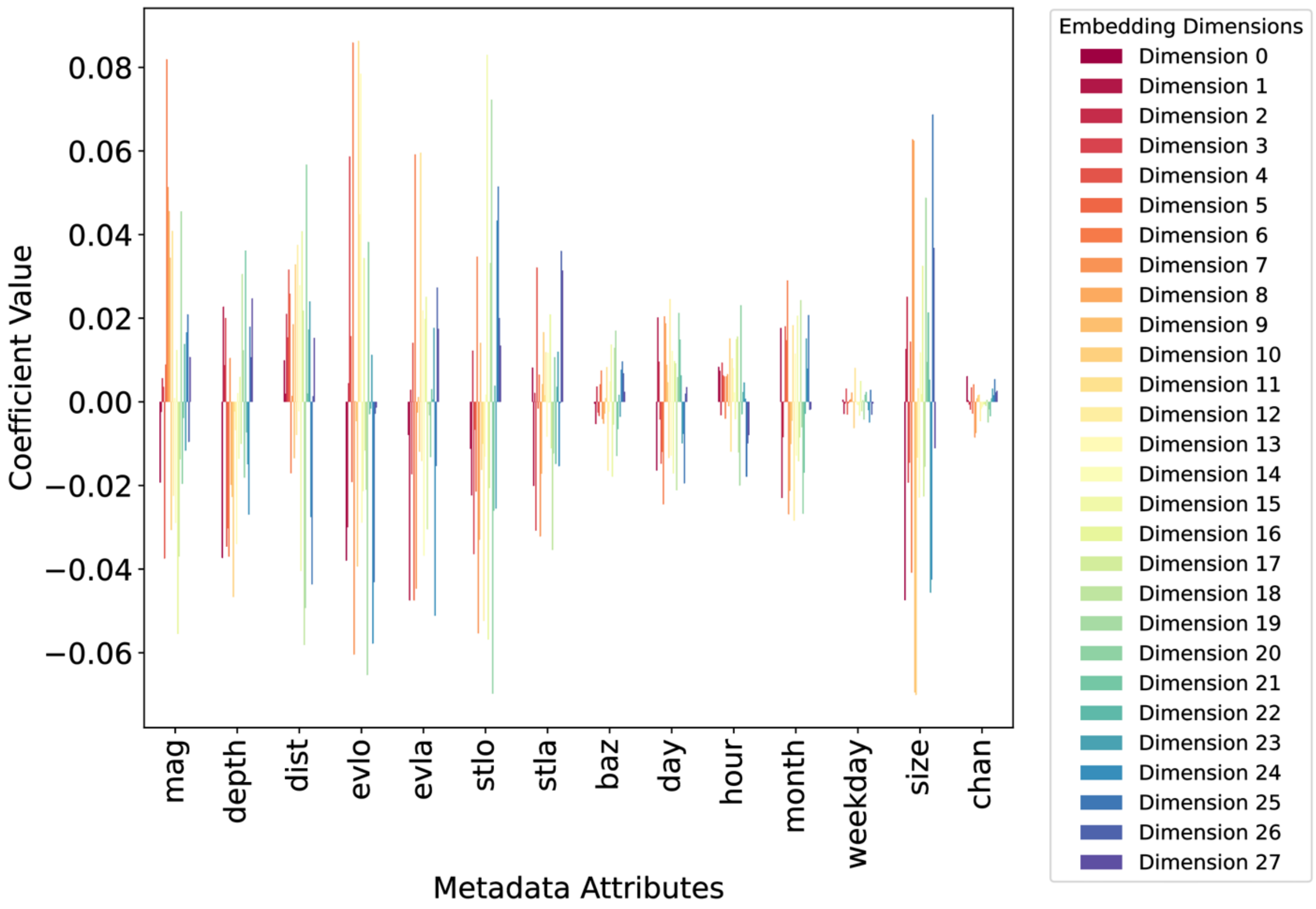
Coefficient Value
0.08
0.06
0.04
0.02
0.00
−0.02
−0.04
−0.06
mag
depth
dist
evlo
evla
stlo
stla
baz
day
hour
month
weekday
size
chan
Metadata Attributes
Embedding Dimensions
Dimension 0
Dimension 1
Dimension 2
Dimension 3
Dimension 4
Dimension 5
Dimension 6
Dimension 7
Dimension 8
Dimension 9
Dimension 10
Dimension 11
Dimension 12
Dimension 13
Dimension 14
Dimension 15
Dimension 16
Dimension 17
Dimension 18
Dimension 19
Dimension 20
Dimension 21
Dimension 22
Dimension 23
Dimension 24
Dimension 25
Dimension 26
Dimension 27

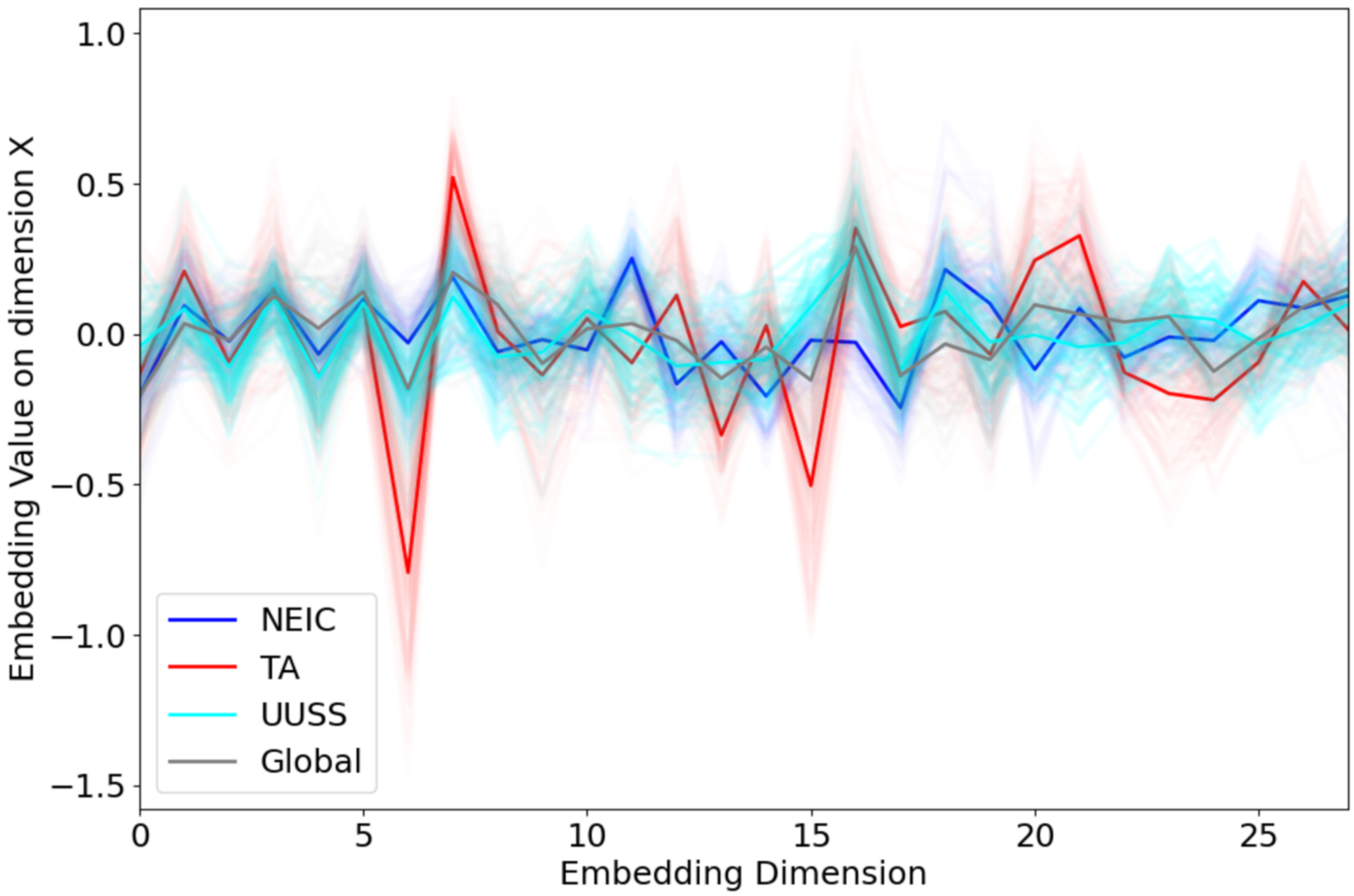

Section XX: Global Cluster exploration

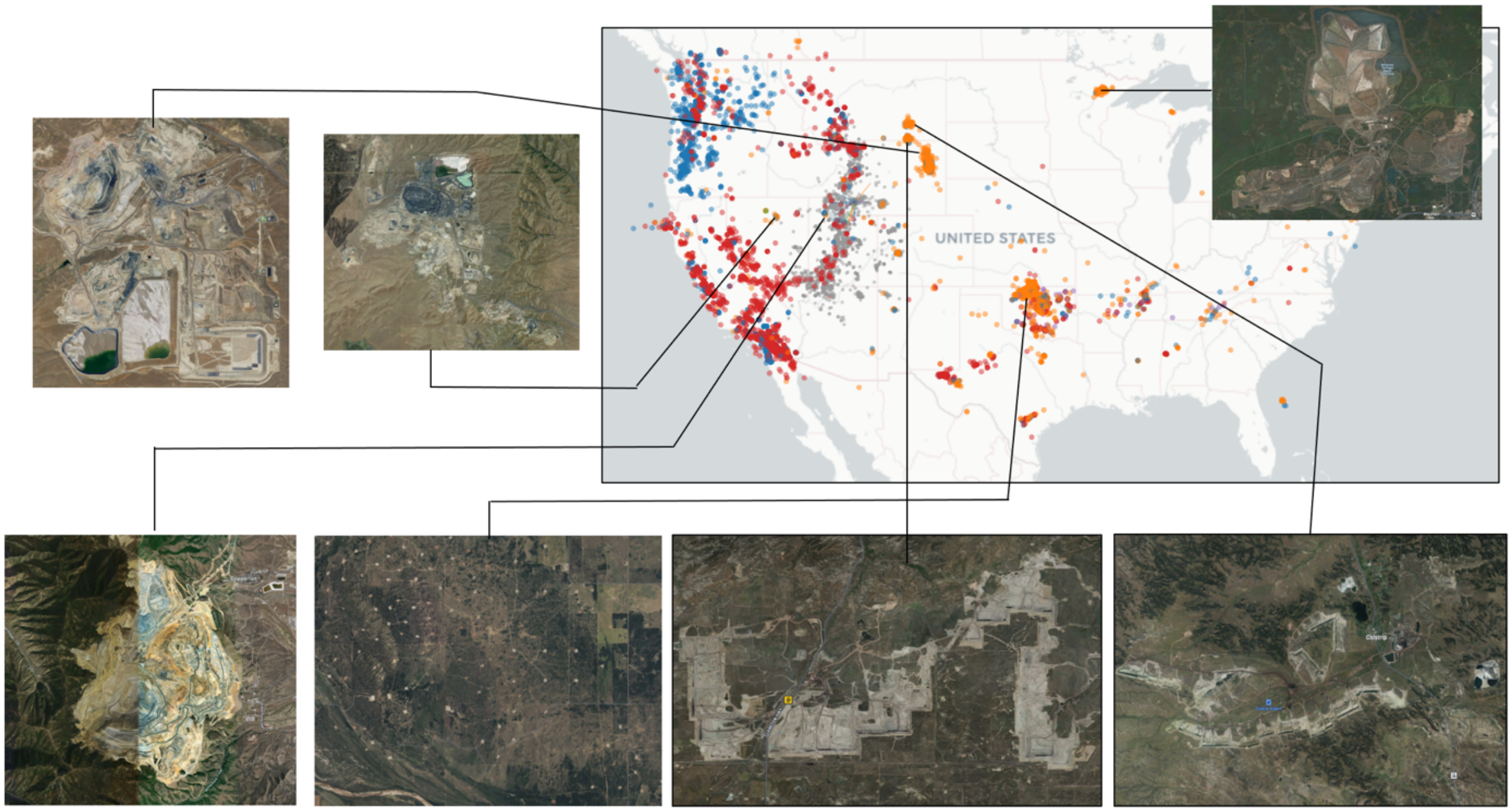

Cascade Mountains
- Rainier
- St Helens

Sawtooth Range

Aleutians

Coso geothermal area

Mammoth-Bishop

Milford Forge Site

Eastern Sierra Hot Springs

Delaware Basin wastewater injection

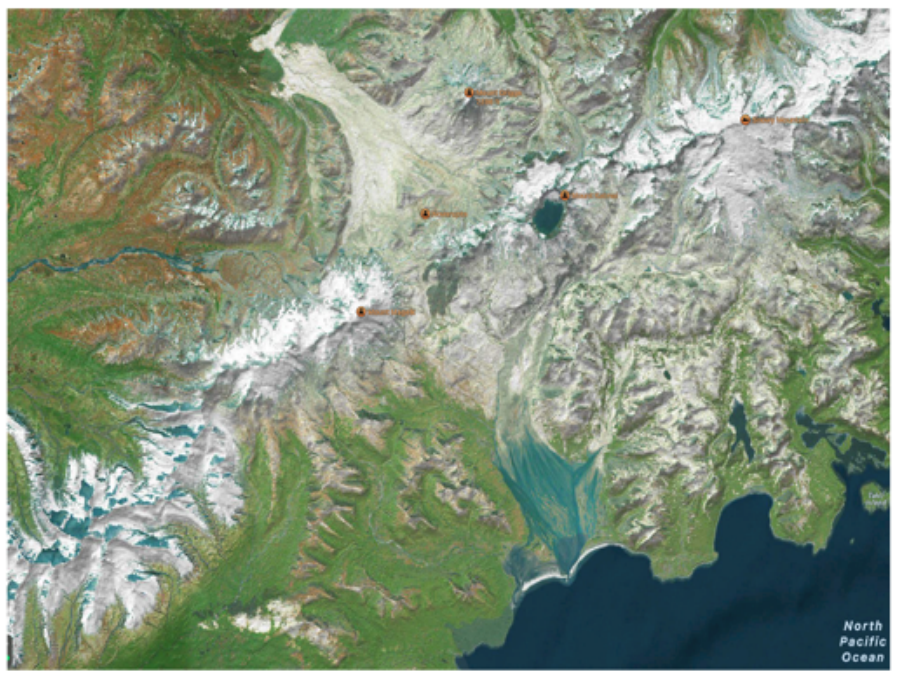

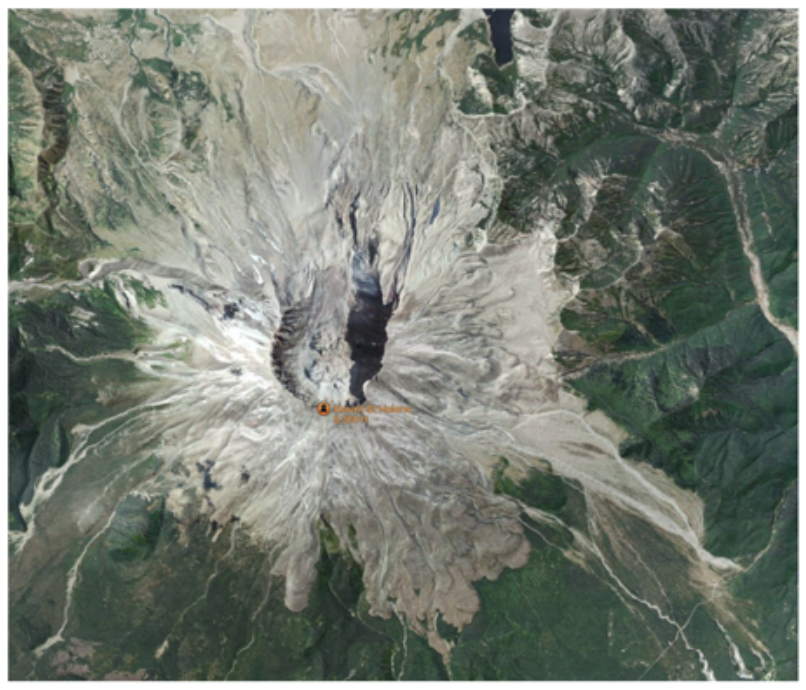
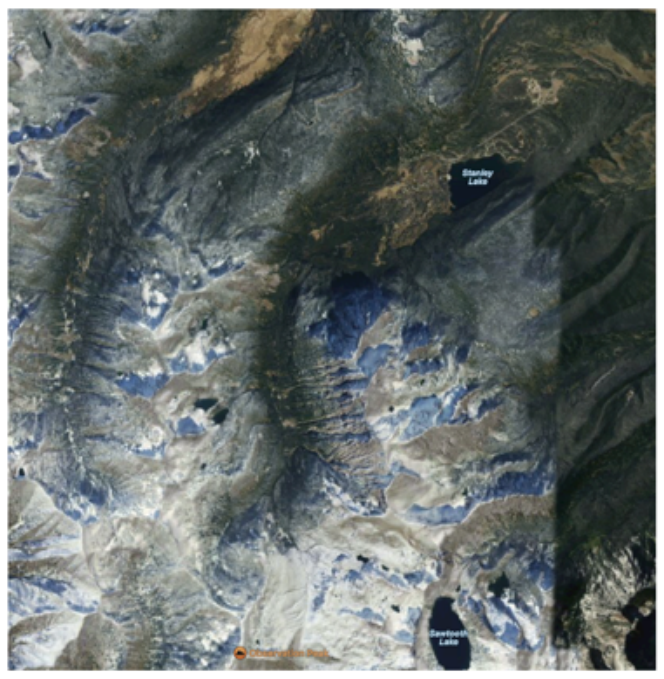
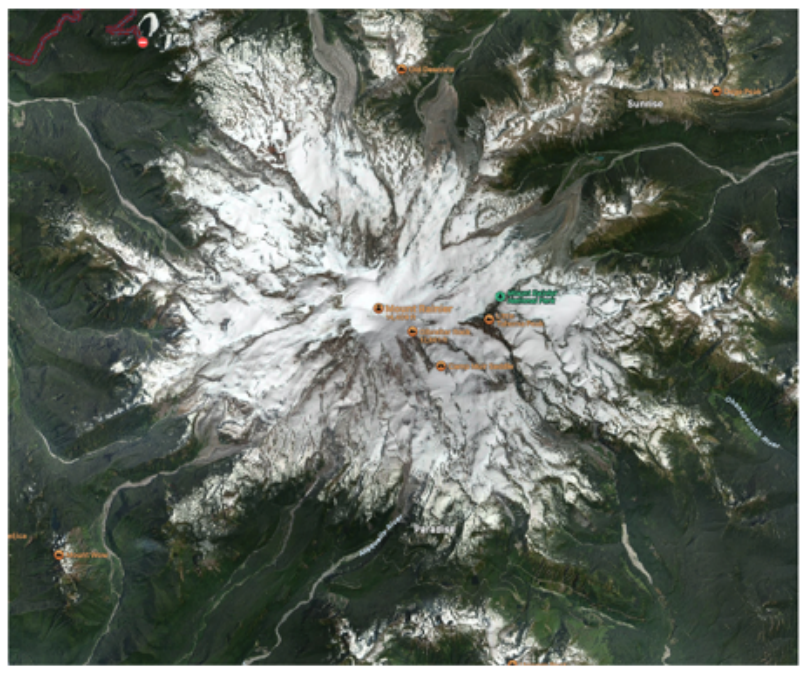

---

They both have a bias towards explostive sources from NEIC, but barlow gets significantly more earthquakes right

Supervised confusion matrix for UUSS:

```
[0.32096018, 0.67903982, 0.        ],
[0.03445075, 0.96554925, 0.        ],
[0.05101198, 0.94898802, 0.        ]
```

Barlow confusion matrix for UUSS on U.S. training:

[0.54783264, 0.45216736, 0.        ],
[0.16599566, 0.83400434, 0.        ],
[0.21933086, 0.78066914, 0.        ]

```
TA: 84
NEIC:92.5
UUSS: 69

For triaxial only
```

[0.62772744, 0.37227256, 0.        ],
[0.22508756, 0.77491244, 0.        ],
[0.32073491, 0.67926509, 0.        ]])

unassigned fraction of training: 0.6176377057721063
unassigned fraction of UUSS: 0.28693848971237745
accuracy on assigned UUSS:
[[0.89249418 0.10750582]
 [0.35706148 0.64293852]] 0.7677163507110807

For UUSS on Global training:

```
[0.54328836, 0.45671164, 0.        ],
[0.12646543, 0.87353457, 0.        ],
[0.1433292 , 0.8566708 , 0.        ]]
```

TA: 78
NEIC: 90
UUSS:71

For triaxial only

```
[[0.77389696, 0.22610304, 0.        ],
  [0.20331708, 0.79668292, 0.        ],
  [0.28346457, 0.71653543, 0.        ]]
```